\documentclass[reprint,superscriptaddress,amssymb,amsmath,aps,prx,longbibliography]{revtex4-2}
\usepackage{minitoc}

\usepackage{graphicx}
\usepackage{dcolumn}
\usepackage{bm}
\usepackage{siunitx}
\usepackage{booktabs}
\usepackage[usenames,dvipsnames]{xcolor}
\usepackage{tcolorbox}
\usepackage{tabularx}
\usepackage{array}
\usepackage{colortbl}
\usepackage{braket}
\usepackage{gensymb}
\usepackage{pdfpages}
\usepackage{amsmath}
\usepackage{mathrsfs}
\usepackage{blindtext}
\usepackage{minitoc}
\usepackage{dsfont}
\usepackage{multirow}
\usepackage{xcolor}
\usepackage{soul}

\usepackage{amssymb}

\makeatletter
\patchcmd{\@outputpage@head}{\@ifx{\LS@rot\@undefined}{}{\LS@rot}}{}{}{}
\makeatother

\tcbuselibrary{skins}

\begin{document}

\title{Quantum Nanophotonic Interface for Tin-Vacancy Centers in Thin-Film Diamond}

\author{Hope Lee}
\thanks{These authors contributed equally to this work.}
\affiliation{Edward L. Ginzton Lab, Stanford University, Stanford, California 94305, United States}

\author{Hannah C. Kleidermacher}
\thanks{These authors contributed equally to this work.}
\affiliation{Edward L. Ginzton Lab, Stanford University, Stanford, California 94305, United States}

\author{Abigail J.M. Stein}
\thanks{These authors contributed equally to this work.}
\affiliation{Edward L. Ginzton Lab, Stanford University, Stanford, California 94305, United States}

\author{Hyunseok Oh}
\affiliation{Department of Physics, University of California, Santa Barbara, California 93106, USA}

\author{Lillian B. Hughes Wyatt}
\affiliation{Materials Department, University of California, Santa Barbara, California 93106, USA}

\author{Casey K. Kim}
\affiliation{Materials Department, University of California, Santa Barbara, California 93106, USA}

\author{Luca Basso}
\affiliation{Sandia National Laboratories, Albuquerque, New Mexico 87123, USA}

\author{Andrew M. Mounce}
\affiliation{Sandia National Laboratories, Albuquerque, New Mexico 87123, USA}

\author{Yongqiang Wang}
\affiliation{Los Alamos National Laboratory, Los Alamos, New Mexico 87545, USA}

\author{Shei S. Su}
\affiliation{Sandia National Laboratories, Albuquerque, New Mexico 87123, USA}

\author{Michael Titze}
\affiliation{Sandia National Laboratories, Albuquerque, New Mexico 87123, USA}
\affiliation{Advanced Instrumentation for Nano-Analytics, Luxembourg Institute of Science and Technology, Belvaux, L-4422 Luxembourg}

\author{Ania C. Bleszynski Jayich}
\affiliation{Department of Physics, University of California, Santa Barbara, California 93106, USA}

\author{Jelena Vučković}
\thanks{jela@stanford.edu}
\affiliation{Edward L. Ginzton Lab, Stanford University, Stanford, California 94305, United States}

\date{\today}


\begin{abstract}
The negatively charged tin-vacancy center in diamond (SnV$^-$) is an excellent solid state qubit with optically-addressable transitions and a long electron spin coherence time at elevated  ($\sim1.7$ K). However, implementing scalable quantum nodes with high-fidelity optical readout of the electron spin state requires efficient photon emission and collection from the system. In this manuscript, we report a quantum photonic interface for SnV$^-$ centers based on one-dimensional photonic crystal cavities fabricated in diamond thin films. \textcolor{black} {Furthermore, we provide a rigorous description of the spontaneous emission dynamics of our system, taking into account individual contributions from both the C and D transitions of the emitter. This allows for determination of Purcell factors per transition and, by extension, the C/D branching ratio SnV$^{-}$ zero phonon line}. We observe quality factors up to $\sim$6000 across this sample, and measure up to a 12-fold lifetime reduction, which translates into a Purcell factor of $F_C=26.2\pm1.5$ for a targeted C transition. \textcolor{black} {By considering the cavity mode polarization alignment with the C and D transition dipole moments, we validate the C/D branching ratio to be $\eta_{\text{BR}}=0.75\pm0.01$, in line with previous theoretical and experimental findings}.
\end{abstract}


\begin{figure*}[ht]
    \includegraphics{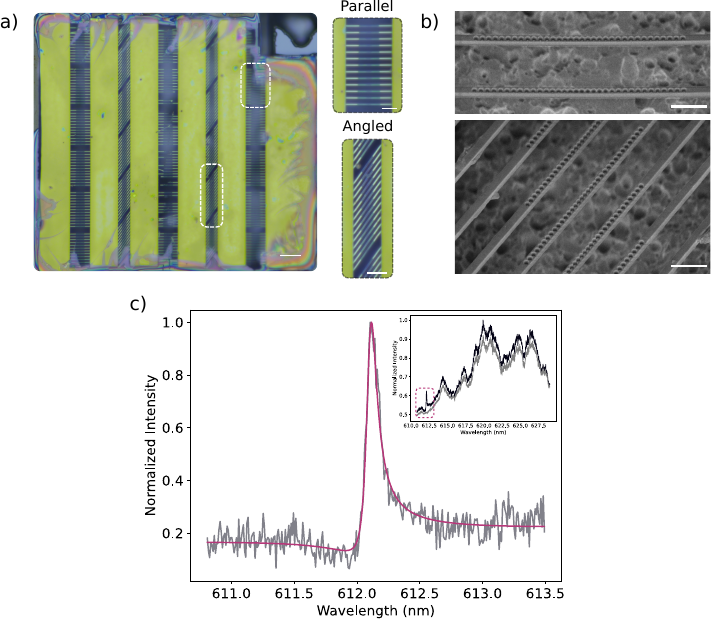}
    \caption{\label{fig:cavity}Cavity design, fabrication, and characterization.
    \textbf{(a)} A white light image of the fabricated devices on the thin film membrane. Representations of the two classes of devices, parallel and angled, are indicated by the white dashed boxes and zoomed to be shown in greater detail. The scale bars indicate 15 $\mu m$ and 5 $\mu m$ for the zoomed out and in images, respectively.
    \textbf{(b)} SEM of fabricated devices. Scale bars indicate $1 \mu$m. 
    \textbf{(c)} The cross polarized reflectivity spectrum for the parallel device. The broadband spectrum is background corrected, then fit to a Fano model, yielding a quality factor of $6032$. The inset shows the broadband reflectivity spectra for both the resonance and background. The fitted region is indicated by the dashed box. \textcolor{black}{In the inset, the background spectrum is represented by the lighter, gray data trace, and the reflectivity spectrum by the darker black trace.}
    }
\end{figure*}

\maketitle

\section{Introduction}\label{sec:intro}

Optically-active defects in semiconductors, or color centers, are a leading platform for quantum network qubits due to their long spin coherence times, bright optical emission, and native compatibility with nanophotonic integration~\cite{Kimble_2008, Awschalom_2018, Atature_2018}. Although the initial keystone quantum network demonstrations utilized the nitrogen vacancy (NV$^{-}$) center in diamond \cite{Pompili_2021, Bernien_2013}, negatively charged group-IV color centers in diamond have garnered interest due to their first-order insensitivity to electric field noise and increased fraction of coherent emission into the zero phonon line (ZPL) \cite{Thiering_2018}. These features render the group-IV vacancy centers significantly more compatible with nanophotonic integration, which is required for downstream scaling of multiple quantum network nodes. The most established of the group-IV vacancies is the negatively charged silicon vacancy center (SiV$^{-}$), which has enabled long-distance networking demonstrations on the scale of those achieved with NV$^{-}$ centers \cite{Knaut_2024}. However, the SiV$^{-}$ suffers from a limited ground state splitting of $\sim50$ GHz, necessitating either mK dilution refrigerator environments or diligent strain engineering for spin state control. In contrast, the rapidly maturing tin-vacancy (SnV$^-$) maintains spin coherence at liquid helium temperatures, around 1.7 K, due to its significantly larger ground state splitting of $\sim$850 GHz \cite{Iwasaki_2017, Rosenthal_2023}.

The tin-vacancy has similarly demonstrated its suitability as a qubit for long distance quantum networks with high fidelity initialization, manipulation, and readout of the qubit spin states and coupling to nanophotonic devices\textcolor{black}{~\cite{Rosenthal_2023, Rosenthal_2024, Rugar_2020, Rugar_2021, Debroux_2021, Guo_2023, Codreanu_2025, Codreanu_2025, Karapatzakis_2024}}. However, photonic interfaces for the SnV$^-$ have thus far been fabricated via diamond bulk carving techniques \cite{Kuruma_2021, Pasini_2024}, severely limiting device performance, fabrication yield, and feasibility of integration with other photonic and electronic components on-chip. Thus the adoption of thin-film diamond paves the way for scalable quantum networks \cite{Ding_2024, Oh_2025, Guo_2021}. 

In this manuscript, we report the fabrication of 1D photonic crystal cavity nanobeams with quality factors of up to $\sim6000$. In particular, we fabricated two orientations of cavity devices, one parallel to and one at $\sim55\degree$ to the $\braket{100}$ axis of the diamond lattice. We measure, for each cavity orientation (`parallel’ and `angled’) the lifetime reduction of two zero phonon line (ZPL) transitions (C and D) of a SnV- color center. These transitions have the same spatial positioning in the cavity field, but orthogonally polarized dipole moments. The ratio of the C to D transition emission rates is referred to as the branching ratio ($\eta_{\text{BR}}$)~\cite{Rugar_2019, Pasini_2024, Thiering_2018}. The conventional figure of merit for color center-cavity coupling is the Purcell factor, defined as the enhancement in spontaneous emission rate for a particular transition when on resonance with the cavity mode. 

\textcolor{black}{For previous reports of SnV$^{-}$ coupling to photonic crystal cavities, Purcell factors were often calculated via applying multiplicative correction factors to the measured lifetime reduction ratio without individual consideration of the C/D transitions and their distinct angular orientations~\cite{Rugar_2021, Kuruma_2021}. To perform a more rigorous treatment, we construct a model to describe the spontaneous emission dynamics of our system and take into account both the C/D transitions, in line with the analysis presented in~\cite{Faraon_2011, Zhang_2018}. Using this model, we report a Purcell factor of up to $F_C=26.2 \pm 1.5$ for the C transition in the angled device, for which the cavity mode polarization and transition dipole moment are best aligned out of the available combinations. Furthermore, we numerically validate the C/D transition branching ratio to be $\eta_{\text{BR}}\approx0.75\pm0.01$, in agreement with previous reports~\cite{Rugar_2019, Pasini_2024, Thiering_2018}, and estimate a global lithographic angular offset of $\delta\psi\approx6.1\degree\pm0.5\degree$. These results constitute crucial progress towards the implementation of scalable quantum networks based on the SnV$^-$.}

\section{Results}\label{sec:results}
\subsection{Device Design and Fabrication}\label{subsec:device_design_fab}

Thin film diamond was prepared via the procedures outlined in \cite{Oh_2025}. Sn$^{2+}$ was implanted into the bulk sample prior to membrane exfoliation with an implantation energy of 380 keV and at a dose of $\num{2e11}$ ions/cm$^2$, targeting SnV$^-$ formation $\sim$90 nm below the surface. Final membrane thicknesses were tuned to 180 nm by reactive ion etching. \textcolor{black}{Further details on thin film preparation are provided in Appendix~\ref{app:thin_film_subsec}.}

Photonic crystal cavities were designed with a 300 nm beam width and 115 nm diameter holes etched into the diamond film. The mirror reflectors consist of 10 holes on each side, while the central cavity consists of 12 holes, with quadratically tapered lattice spacings. In order to account for fabrication infidelities and ensure resonances in proximity of the SnV$^{-}$ 619 nm ZPL wavelength, we sweep the lattice spacings from 180 nm to 210 nm in steps of 2.5 nm per device.

For the cavities reported in this manuscript, we measure a post-fabrication beam width of $326 \pm 13$ nm, average hole diameter of $134 \pm 3$ nm, and lattice constant of 210 nm for the parallel device; and a beam width of $325 \pm 2$ nm, average hole diameter of $119 \pm 4$ nm, and lattice constant of 197.5 nm for the angled device, determined via \textcolor{black}{scanning electron microscopy (SEM)} (Fig.~\ref{fig:cavity}b). We note that our etch recipe renders a $\sim5-10\degree$ sidewall angle. Taking these dimensions into account, we simulate quality factors (Q) of $2.8 \times 10^5$ ($2.3 \times 10^4$) and mode volume (V) 0.46 (0.46) $(\frac{\lambda}{n})^3$ for the parallel (angled) device. \textcolor{black}{Further discussion of device design and fabrication yield is provided in Appendices ~\ref{app:cavity_fab_subsec} and ~\ref{app:cavity_yield}.} 
\textcolor{black}{In order to optimize space usage of the limited diamond film, transmission access to the cavities (e.g. via grating couplers or adiabatic tapers) was forgone and instead all devices were probed confocally out of plane.} 

\subsection{Characterization of Devices and Color Centers}\label{subsec:characterization}

Throughout measurements, the sample is maintained at 4 K in a Montana Instruments closed cycle cryostat. We utilize two discrete optical paths, subsequently referred to as the i) `cross-polarized' and ii) photoluminescence (`PL') paths. The two optical paths enable independent access to the cavity resonance and color center signals. 

We start by probing our cavity resonances using the cross-polarized reflectivity path. The cavities are mounted such that the resonance modes for parallel devices are primarily vertically polarized. \textcolor{black}{Devices are excited confocally via a broadband supercontinuum laser, and the reflected signal is collected confocally and detected by a spectrometer. The excitation and collection paths are mixed via a polarizing beam splitter (PBS), and a half-wave plate (HWP) is inserted in the shared path right before the cryostat. The HWP is rotated to either $\approx45\degree$ or $\approx100\degree$ to optimize the resonance SNR of the parallel devices or angled devices, respectively.} Further detail on the optical setup is provided in Appendix~\ref{app:measurement_setup}. 

Each broadband reflectivity spectrum is background-corrected and then fit to a Fano model. For the parallel device, we determine a quality factor of $6032$ (Fig.~\ref{fig:cavity}c). The quality factor of the angled device was determined to be $3942$, as discussed in Appendix~\ref{app:angled_cavity} (Fig.~\ref{sfig:angled_cavity_01}). \textcolor{black}{We note that this method of fitting a Fano model to the spectral features in reflectivity serves as a lower bound estimator on the quality factors. Further discussion is provided in Appendix~\ref{app:Q_determination}.}

\begin{figure*}
    \includegraphics{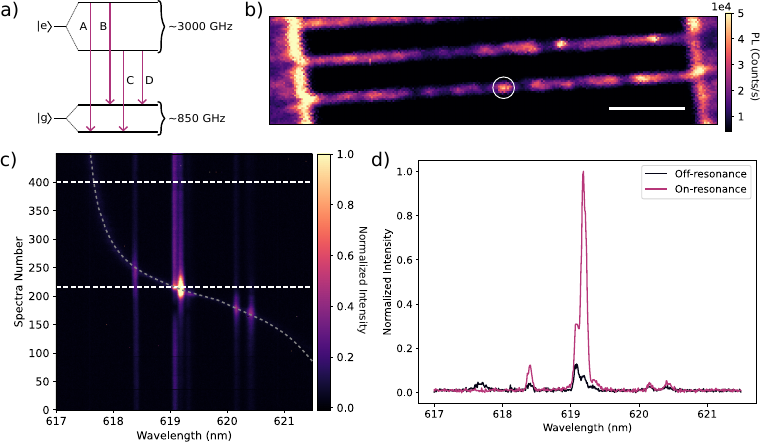}
    \caption{\label{fig:PL_enhancement}SnV$^{-}$ level structure, PL confocal scan and cavity enhancement.
    \textbf{(a)} Schematic of the orbital energy states of the SnV$^{-}$. Characteristic of a group-IV color center, the ground ($\ket{g}$) and exited state ($\ket{e}$) are split via the combined effects of spin-orbit coupling and the Jahn-Teller effect. For the SnV$^-$ the ground state splitting is $\sim850$ GHz, and the excited state splitting $\sim3000$ GHz; these state splittings yield four separate ZPL transitions. In cryogenic conditions, PL signal is dominated by the two longer wavelength, lower energy transitions, labeled as C and D~\cite{Görlitz_2020}. 
    \textbf{(b)} PL confocal scan of the parallel device. The emitter cluster addressed is indicated by the white circle. The scale bar indicates $3 \mu$m.
    \textbf{(c)} PL enhancement via gas tuning. The cavity resonance is first red-shifted via Ar gas condensation past all SnV$^-$ transitions of interest. The sample is then naturally `back-tuned' over the course of repeated PL \textcolor{black}{spectra}. The two spectra of interest are indicated by the white dashed lines. The gray dashed line is an approximate guide for the eye of the cavity resonance.
    \textbf{(d)} PL spectra on and off resonance with the cavity. We see that the most strongly enhanced SnV$^-$ transition demonstrates a $\sim$10-fold PL enhancement.
    }
\end{figure*}

\subsubsection{PL enhancement} \label{subsec:pl_enhancement}
Next, we perform photoluminescence (PL) spectroscopy to identify emitters located in the cavity mode, exciting with $\sim$5 mW of a 520 nm CW source and filtering collected counts with a 620/14 nm bandpass filter (Fig.~\ref{fig:PL_enhancement}b). Due to the high implantation density, approximately 2-3 emitters lie within our cavity mode. To probe for emitter-cavity coupling, we first red-shift the cavity resonance past the identified SnV$^-$ transitions via argon gas condensation. Then, we `back-tune' the resonance controllably by illuminating the nanobeam with $\sim$0.75 mW of green excitation power to evaporate condensed \textcolor{black}{gas}, while simultaneously collecting PL spectra to characterize SnV$^-$ PL enhancement (Fig.~\ref{fig:PL_enhancement}c). For the parallel device, we identify a $\sim$10-fold PL enhancement of the emitter when the cavity is tuned into resonance with the transition (Fig.~\ref{fig:PL_enhancement}d). This measure of PL enhancement indicates significant emitter-cavity coupling, but does not suffice as an accurate measure of the Purcell factor given uncertainties in exact collection efficiencies. PL enhancement measurements for the angled device are provided in Appendix~\ref{app:angled_cavity}.


\subsubsection{Lifetime reduction measurement and calculation}
\label{subsec:lifetime_reduction}
To quantify Purcell factors, we measure the lifetime reduction of the emitter when the cavity is tuned into resonance. As before, we red-detune the cavity resonance beyond the SnV$^-$ transition in preparation to back-tune across the target wavelength. However, we now toggle our excitation between the cross polarized path, used to both monitor and back-tune the resonance wavelength, and the PL path, used to measure the emitter optical lifetime. In the cross polarized path, we excite simultaneously with both the supercontinuum and CW green diode laser (set again to $\sim$0.75 mW) in order to both monitor and backtune the cavity resonance. In the PL path, we isolate a single transition with a narrow band ($\sim$0.3 nm) tunable filter, and excite the emitter above band with a 520 nm pulsed laser. We apply excitation pulses with 16 ns pulse widths, a 3.3 MHz repetition rate, and an averaged power of 270 $\mu$W. Each lifetime trace is integrated for 3 minutes to build up suitable count rates, while each cross-polarized resonance spectrum is integrated for 3 seconds to allow ample time for the cavity resonance to gradually back-tune across the transition.

\begin{figure*}
     \includegraphics[width=0.7\textwidth]{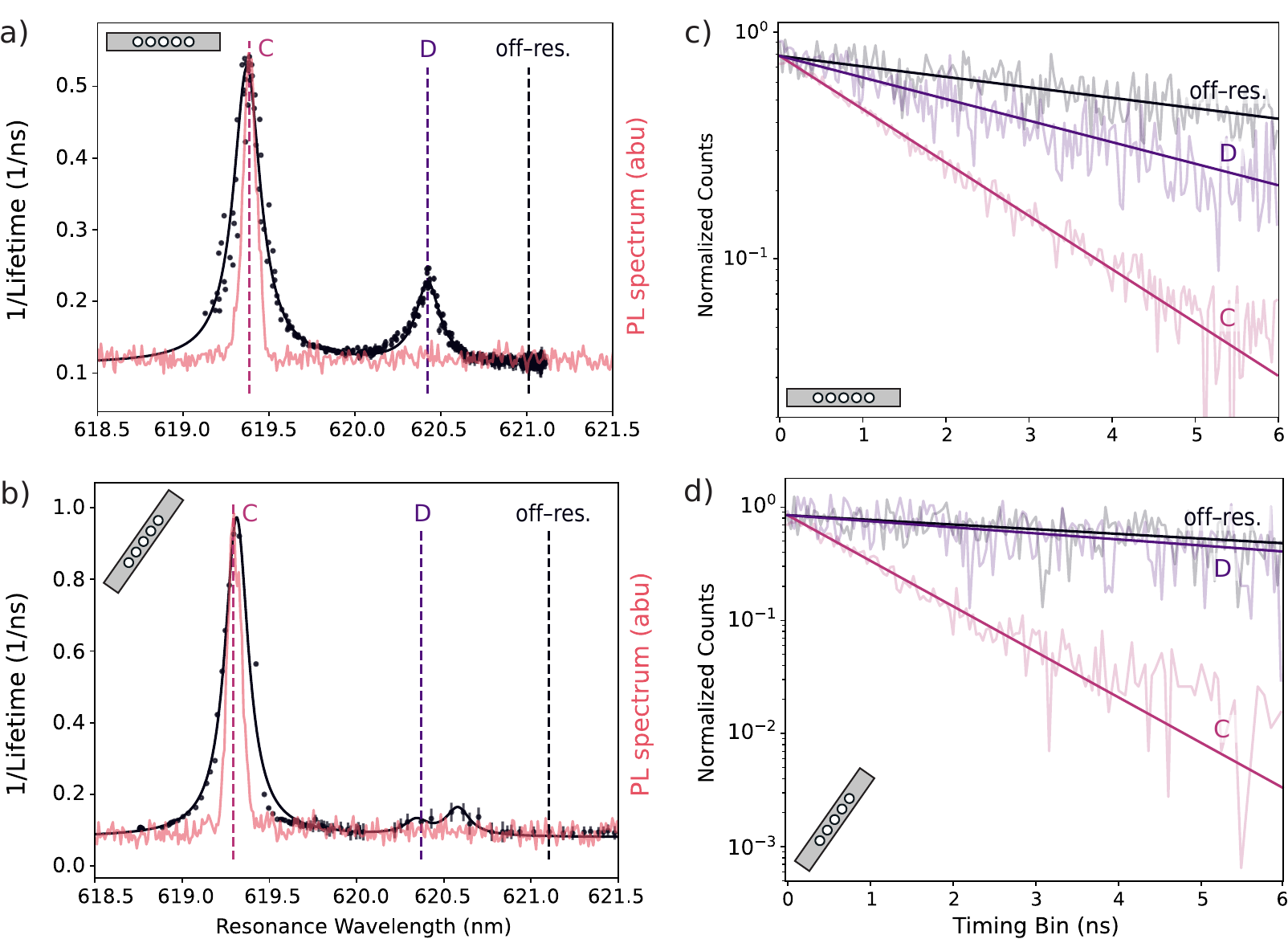}
    \caption{\label{fig:lifetime_reduction}
    Lifetime reduction and branching ratio analysis.
    \textbf{(a)} Spontaneous emission rate vs cavity resonance wavelength for the parallel device (black). The data are fit to \textcolor{black}{a model consisting of the sum of two Lorentzians}. The PL spectrum (red) is overlayed. We extract from this fit model the amplitude ratios $\zeta_C=4.672$ and $\zeta_D=1.985$. The three vertical dashed lines indicate the lifetime slices that are plotted separately, representing the traces with the C transition on resonance (pink), D transition on resonance (purple), and both transitions off resonance (black).
    \textbf{(b)} Spontaneous emission rate vs cavity resonance wavelength for the angled device (black). As in (a), the PL spectrum (red) is overlayed. For this device, we extract from this fit model the amplitude ratios $\zeta_C=12.230$ and $\zeta_D=1.514$. We note that due to filtering limitations, there are contributions from two separate emitters, and the data are thus fit to \textcolor{black}{the sum of four Lorentzians}. The dashed lines indicate the lifetime slices of interest, as in panel (a).
    \textbf{(c)} Normalized lifetimes of the SnV$^-$ with C/D transitions on and off resonance for the parallel device. We measure an off-resonance lifetime of $9.412 \pm 0.09$ ns. When the C(D) transition is on-resonance with the cavity, the lifetime is reduced to $1.847 \pm 0.002$ ns ($4.570 \pm 0.02$ ns). The lifetimes are fit to amplitude normalized and background corrected single exponential models.
    \textbf{(d)} Normalized lifetimes of the SnV$^-$ with C/D transitions on and off resonance for the angled device. We measure an off-resonance lifetime of $10.507 \pm0.2$ ns. When the C(D) transition is on-resonance, the lifetime is reduced to $1.079 \pm 0.002$ ns ($8.109 \pm 0.2$ ns). The lifetimes are fit to amplitude normalized and background corrected single exponential models. Due to the contributions of a second emitter transition, at longer timescales, the data begins to deviate slightly from a single exponential \textcolor{black}{model}. However, within the fitting range, the observed fluorescence lifetime is primarily dominated by a single decay timescale.
    }
\end{figure*}

We target selective collection from the C transition, but due to \textcolor{black}{the C/D transitions sharing an excited state}, we see significant lifetime reduction when the cavity is on resonance with \textcolor{black}{either transition individually}. Furthermore, we will show in Section~\ref{subsec:purcell} that by considering the orthogonality of the C and D dipoles, we can extract individual Purcell factors for each transition, \textcolor{black}{and by extension, the C/D branching ratio} . In preparation for this analysis, we calculate the spontaneous emission rate from the measured emitter decay lifetime via $\Gamma=1/\tau_{\text{lifetime}}$. When resonant with the cavity, transition C(D) demonstrates a lifetime of \textcolor{black}{$1.85 \pm 0.02$ ns}, or emission rate $0.54$ 1/ns (\textcolor{black}{$4.57 \pm 0.02$ ns}, or emission rate of $0.22$ 1/ns) compared to the off-resonance lifetime of $9.41 \pm 0.09$ ns, or emission rate of $0.11$ 1/ns (Fig.~\ref{fig:lifetime_reduction}c). We then fit the data to the \textcolor{black}{sum of two Lorentzians}. Dividing a fitted Lorentzian peak to the background \textcolor{black}{yields an} emission rate enhancement ratio, which we denote as $\zeta$. For the C and D transitions, we extract \textcolor{black}{$\zeta_C=4.67\pm0.07$ and $\zeta_D=1.99\pm0.02$}, respectively (Fig.~\ref{fig:lifetime_reduction}a).

For the angled device, there remain contributions from a secondary emitter in the collected counts. We therefore fit the spontaneous emission rates to \textcolor{black}{the sum of four Lorentzians but focus our analysis on the more prominent emitter}. We determine that transition C(D) demonstrates a lifetime of \textcolor{black}{$1.08 \pm 0.01$ ns} , or emission rate $0.93$ 1/ns (\textcolor{black}{$8.11 \pm 0.23$ ns}, or emission rate $0.12$ 1/ns) compared to the off-resonance lifetime of \textcolor{black}{$10.51 \pm 0.19$ ns}, or emission rate $0.10$ 1/ns (Fig.~\ref{fig:lifetime_reduction}d). The fitted amplitude ratios are \textcolor{black}{$\zeta_C=12.23\pm0.69$ and $\zeta_D=1.51\pm0.27$} (Fig.~\ref{fig:lifetime_reduction}b).

\subsubsection{Purcell factor analysis}
\label{subsec:purcell}

\textcolor{black}{Although our time-resolved fluorescence measurements are spectrally filtered to collect signal only from the C transition, the observed spontaneous emission rate reflects the dynamics of the excited-state population. The excited-state population can decay by either the C/D ZPL transitions or the PSB, and, consequently, the C transition emission rate is enhanced when the cavity is tuned in resonance with either transition. This therefore motivates a rate-equation treatment in order to precisely extract the Purcell enhancement of each individual ZPL transition, consistent in construction with that presented in~\cite{Faraon_2011, Zhang_2018}.}



We express the spontaneous emission dynamics of the full cavity-emitter system with the following expressions for spontaneous emission rates:

\begin{align}\label{eq:rates}
    \Gamma_\text{0} &= \Gamma_C + \Gamma_D + \gamma_{\text{PSB}} \\
    \Gamma_\text{coupled} &= F_C*\Gamma_C + F_D*\Gamma_D + \gamma_{\text{PSB}},
\end{align}

\noindent where $\Gamma_\text{0}$ is the uncoupled or bulk emission rate, and $\Gamma_\text{coupled}$ is the cavity-coupled enhanced spontaneous emission rate. In our analysis, we neglect contributions from the A/B ZPL transitions of the SnV$^{-}$, as the two higher energy transitions are suppressed in cryogenic conditions~\cite{Iwasaki_2017}. Therefore, in the uncoupled case, the total spontaneous emission rate is given by the sum of $\Gamma_C$, the C transition emission rate, $\Gamma_D$, the D transition emission rate, and $\gamma_{\text{PSB}}$, the rate of emission into the phonon side band (PSB). In the cavity-coupled case, $F_C$ and $F_D$ represent the Purcell factor that enhances the emission into the two orthogonal ZPL transitions.  

Dividing $\Gamma_\text{coupled}$ by $\Gamma_0$ to analyze the emission enhancement ratios, we write:

\begin{equation}\label{eq:ratio}
    \begin{aligned}
        \zeta = \frac{\Gamma_\text{coupled}}{\Gamma_0} &= F_C*\frac{\Gamma_C}{\Gamma_0} + F_D* \frac{\Gamma_D}{\Gamma_0} + \frac{\gamma_{\text{PSB}}}{\Gamma_0} \\
            &= F_C* \eta_{\text{DW}}\eta_{\text{BR}}\\&\phantom{=}+F_D* \eta_{\text{DW}}(1-\eta_{\text{BR}}) +(1-\eta_{\text{DW}}) 
    \end{aligned}
\end{equation}

\noindent where $\eta_{\text{DW}} = 0.57$ is the Debye-Waller factor, or coherent ZPL proportion of total radiative emission~\cite{Görlitz_2020}. $\eta_{\text{BR}}$ is the branching ratio between transitions C and D. We are able to write two separate ratios, $\zeta_C$ and $\zeta_D$, for when the cavity is on resonance with transition C and D, respectively. For ratio $\zeta_C$ ($\zeta_D$), we can set $F_D$ ($F_C$) equal to $1$.

We now recast these equations to solve for the Purcell factors:


\textcolor{black}{\begin{align}\label{eq:purcell}
F_C &=1 + \frac{\zeta_C - 1}{\eta_{\text{DW}}\eta_{\text{BR}}}\\
F_D &=1 + \frac{\zeta_D - 1}{\eta_{\text{DW}}(1-\eta_{\text{BR}})}
\end{align}}


\textcolor{black}{By formulating transition-specific rate-equations, we avoid relying on `lump-sum' multiplicative correction factors or omitting explicit treatment of the individual C and D transitions ~\cite{Rugar_2021, Kuruma_2021}.}

\textcolor{black}{Using the literature branching ratio value $\eta_\text{BR}=0.78$ and the expressions derived above,} we determine Purcell factors of \textcolor{black}{$F_C=26.2 \pm 1.5$} and \textcolor{black}{$F_D=5.1 \pm 2.2$} (\textcolor{black}{$F_C=9.2 \pm 0.2$} and \textcolor{black}{$F_D=8.9 \pm 0.1$}) for the angled (parallel) cavity~\cite{Thiering_2018, Rugar_2019, Pasini_2024}. For the second emitter of the angled device, we can estimate Purcell factors of \textcolor{black}{$F_C=1.1 \pm 0.4$} and \textcolor{black}{$F_D=8.8 \pm 2.2$}. We note that since the second emitter in the angled cavity demonstrates a larger Purcell factor for the D transition, we conclude that the emitter is oriented orthogonally to the other two emitters of study. Additional discussion regarding the second emitter is provided in Appendix~\ref{app:angled_cavity_2nd_emitter}.

\begin{figure*}
    \includegraphics[width=\textwidth]{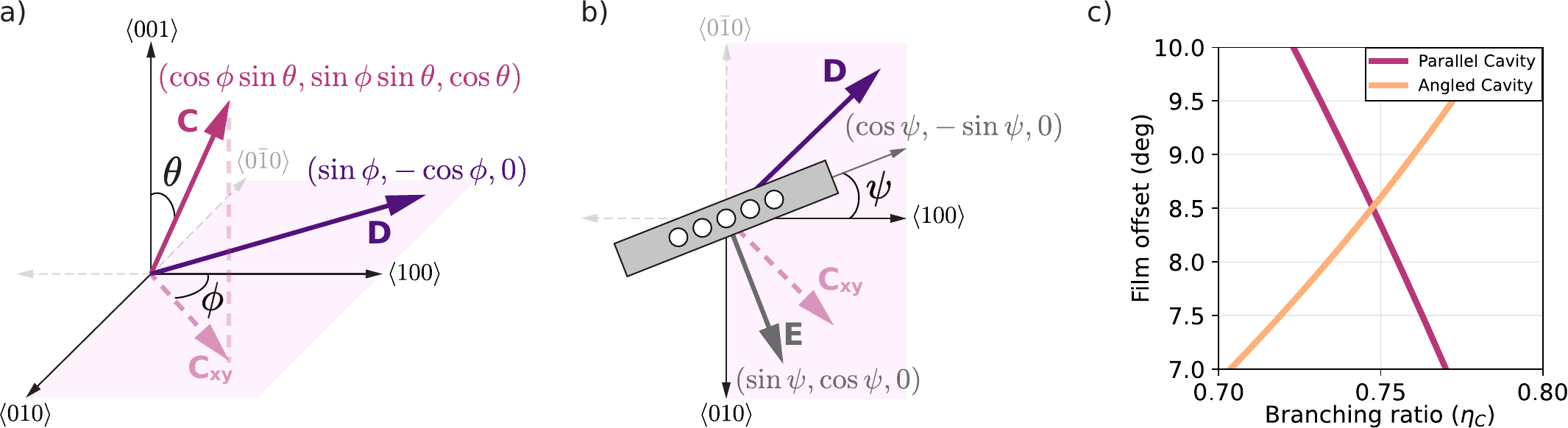}
    \caption{\label{fig:angular_analysis}\textcolor{black}{
    Cavity mode polarization and SnV$^{-}$ transition dipole moment alignment.
    \textbf{(a)} Dipole orientations for a $\langle100\rangle$-oriented diamond. The C dipole points along lattice vector $\langle111\rangle$, giving a fixed out-of-plane tilt $\theta$, and an in-plane projection C$_{xy}$, with azimuthal angle $\psi$. The D dipole lies in-plane along lattice vector $\langle1\bar{1}0\rangle$. For the purposes of the analysis, these dipole moment unit vectors are translated into real-space and mapped onto the sample-fixed axes $\hat{x}\parallel\langle100\rangle$, $\hat{y}\parallel\langle010\rangle$, and $\hat{z}\parallel\langle001\rangle$. The pink region denotes the sample surface plane.
    \textbf{(b)} Schematic of the sample surface and cavity orientation. The cavity is fabricated at an angle $\psi$ relative to $\langle100\rangle$. The fundamental nanobeam mode polarization $\mathbf{E}$ is taken to be orthogonal to the cavity axis. The resulting polarization overlap with the transition dipole moments determines the Purcell enhancement of the C and D transitions. The pink region denotes the sample surface plane.
    \textbf{(c)} Solution of the analysis in Section~\ref{subsec:lifetime_reduction} in allowing both $\eta_{\text{BR}}$ and $\delta\phi$ to vary freely. We determine a branching ratio $\eta_\text{BR}\approx0.75\pm0.01$ and fabrication offset $\delta\psi\approx8.51\degree$.}}
\end{figure*}

\subsubsection{\textcolor{black}{Cavity mode polarization and emitter dipole moment angular alignment}}
\label{sec:cavity_dipole_angle}

\textcolor{black}{We now extend the presented analysis to extracting the C/D branching ratio and cavity angular orientation and consider the relative orientation between the cavity mode polarization and the C/D transition dipole moments. The dependence of the Purcell factor on the angular alignment between the emitter dipole moment and the cavity field is given by:}

\begin{equation}
\textcolor{black}{
\label{eq:purcell_ideal}
F = F_\text{o}
\;\frac{\bigl|\Vec{\boldsymbol{\mu}}\cdot \Vec{\mathbf{E}}(\mathbf{r})\bigr|^2}{\|\Vec{\boldsymbol{\mu}}\|^2\,\|\Vec{\mathbf{E}}(\mathbf{r})\|^2}}
\end{equation}

\textcolor{black}{Where $\Vec{\boldsymbol{\mu}}$ is the dipole moment the quantum emitter transition, $\Vec{\mathbf{E}}\mathbf{(r)}$ is the electric field at the position of the emitter, and $F_{o}$ is comprised of all other contributions required to determine the Purcell factor (Q/V ratio, emitter spatial position, and spectral overlap between the emitter transition and cavity resonance wavelengths). In the case of a single SnV$^{-}$ center, $F_{o}$ contains all multiplicative factors that are shared for the respective Purcell factors of the C and D transitions---the singular difference for the two transitions stems from the dipole moment-cavity field alignment. As such, we express $F_C$ and $F_D$ in the form:}

\textcolor{black}{
\begin{equation}
\label{eq:purcell_CD_def}
\begin{aligned}
F_C &= F_o\;\bigl|\mathbf{\hat C}\cdot \hat{\mathbf{e}}_\mathrm{cav}\bigr|^2\\
F_D &= F_o\;\bigl|\mathbf{\hat D}\cdot \hat{\mathbf{e}}_\mathrm{cav}\bigr|^2
\end{aligned}
\end{equation}}

\textcolor{black}{where $\mathbf{\hat C}$($\mathbf{\hat D}$) is the unit dipole moment vector of the C(D) transition of a single SnV$^{-}$ center and $\hat{\mathbf{e}}_\mathrm{cav}$ is the unit polarization vector of the cavity mode. In a $\langle 100\rangle$-oriented diamond, the $C$ dipole possesses a substantial out-of-plane component, whereas the $D$ dipole lies entirely in the sample plane (Fig.~\ref{fig:angular_analysis})~\cite{Hepp_2014}. This motivates the development of $\langle 111 \rangle$ diamond growth and devices, where crystallographic orientation is engineered to enable more favorable emitter--cavity alignment~\cite{Codreanu_2025}.}

\textcolor{black}{
For the purposes of our analysis, we require writing out all relevant vectors in real-space coordinates. We therefore map the crystallographic dipole directions to real space as $\hat{x}\parallel\langle100\rangle$, $\hat{y}\parallel\langle010\rangle$, and $\hat{z}\parallel\langle001\rangle$ and write:}


\textcolor{black}{\begin{equation}
\label{eq:dipoles_defined}
\begin{aligned}
\hat{\mathbf{C}}
&= \left\langle \cos\phi\sin\theta,\ \sin\phi\sin\theta,\ \cos\theta \right\rangle \\
\hat{\mathbf{D}}
&=\left\langle \sin\phi,\ -\cos\phi,\ 0 \right\rangle\\
\hat{\mathbf{e}}_\mathrm{cav} &= \langle \sin\psi,\ \cos \psi,\ 0\rangle
\end{aligned}
\end{equation}}

\textcolor{black}{where $\phi$ and $\psi$ are azimuthal angles and $\theta$ is a polar angle. For the SnV$^{-}$ in $\langle100\rangle$ diamond, $\phi\approx45\degree$ and $\theta\approx55\degree$. $\hat D$ is oriented orthogonal to $\hat C$ in projection to the $xy$ plane, and therefore the effective azimuthal angle is $\phi'=\phi-90\degree$. The expression for $\hat D$ as written in Eq.~\ref{eq:dipoles_defined} is simplified by trigonometric identifies. Crystallographic miscut of the diamond during polishing ($\approx1\degree$) has been neglected in this analysis. $\psi$ represents the angle of the cavity nanobeam, which is lithographically defined during fabrication. In our case, we aimed for $\psi\approx0\degree$ for parallel and $\psi\approx55\degree$ for angled devices. $\hat{\mathbf{e}}_\mathrm{cav}$ has an effective azimuthal angle of $\psi'=\psi+90\degree$ as the electric field polarization of the cavity mode is orthogonal to the nanobeam orientation; the presented unit vector is simplified via trigonometric relations. A visual representation of the dipoles and cavity vectors is provided in Figure \ref{fig:angular_analysis}.}

\textcolor{black}{Substituting Eq.~\eqref{eq:dipoles_defined} into Eq.~\eqref{eq:purcell_CD_def} yields:}

\textcolor{black}{\begin{align}
F_C
&= F_o\;\tfrac{1}{2}\sin^255\degree\;\bigl(1+\sin 2\psi\bigr),
\label{eq:theoretical_purcell_Cpsi}\\
F_D
&= F_o\;\tfrac{1}{2}\bigl(1-\sin 2\psi\bigr).
\label{eq:theoretical_purcell_Dpsi}
\end{align}}

\textcolor{black}{Taking the ratio eliminates $F_o$:
\begin{equation}
\frac{F_D}{F_C}
= \frac{1}{\sin^255\degree}\;
\frac{1-\sin 2\psi}{1+\sin 2\psi} \label{eq:FD_over_FC}
\end{equation}}

\textcolor{black}{Thus, the experimentally extracted Purcell factors provides a direct estimate of the cavity orientation $\psi$. Using the experimentally determined values for $F_C$ and $F_D$ as reported in Subsection~\ref{subsec:purcell}, Eq.~\ref{eq:FD_over_FC} yields:}

\textcolor{black}{\begin{equation}
\psi_\mathrm{parallel} = 6.1\pm0.5\degree,\quad
\psi_\mathrm{angled}   = 64.9\pm27.6\degree.
\end{equation}
From these values, we can calculate the difference between the two angles $|\psi_\mathrm{parallel}-\psi_\mathrm{angled}|\approx58.8\degree$, which exhibits a slight discrepancy from the expected $55\degree$, as was lithographically defined. However, we note that the error for $\psi_\mathrm{angled}$ is considerable, given the large uncertainty in fitting the weakly enhanced $D$ transition of the angled cavity.}

\textcolor{black}{Additionally, we can interpret $\psi_\mathrm{parallel}$ as a global lithographic angular offset, $\delta\psi$. By allowing both $\delta\psi$ and branching ratio $\eta_\text{BR}$ to vary as free parameters, the two values can be jointly fit. Performing this analysis yields $\delta\psi \approx 8.51\degree$ and $\eta_{\text{BR}} \approx 0.75\pm0.01$, the latter notably being consistent with literature values~\cite{Rugar_2019, Pasini_2024, Thiering_2018}.}


\textcolor{black}{As expected, for $\langle 100\rangle$ diamond, angling photonic crystal cavities such that $\psi = 45^\circ$ would yield maximal overlap between cavity mode polarization and the in-plane projection dipole moment of either the C or D transition. Although the D transition dipole moment can nominally be perfectly aligned in-plane for $\langle100\rangle$ diamond substrates, its smaller branching ratio renders optimizing coupling to the C transition favorable~\cite{Codreanu_2025}.}

\section{Discussion}\label{sec:discussion}
Our results highlight the importance of analyzing the lifetime reduction for both the C and D transitions. By explicitly modeling the shared excited-state spontaneous emission dynamics and accounting for the orthogonal polarizations of the C and D transition dipole moments, we are able to extract not only the individual Purcell factors of each transition, but also the intrinsic C/D branching ratio of the SnV$^-$. 

This contrasts with prior work where Purcell factors are reported per SnV$^-$ center and calculated by applying \textcolor{black}{multiplicative correction factors without writing out an explicit, full treatment of the spontaneous emission dyanmics of the system} \cite{Rugar_2021, Kuruma_2021}. Furthermore, \textcolor{black}{experimental determination of branching ratios previously relied on PL spectra or determined through quasi-resonant excitation} \cite{Rugar_2019, Zhang_2018}. Distinguishing the Purcell enhancements of the C and D transitions provides a more faithful description of the underlying physics, which indicates the optimal emitter-cavity alignment for maximizing subsequent spin-state readout fidelity. 

The asymmetry in cavity coupling between orthogonal transitions further serves as a sensitive probe of dipole orientation relative to the cavity field, allowing us to extract the angular alignment of the fabricated devices with respect to the crystal axes. If the collective lithography offset from the main lattice vector is known independently, for example via \textcolor{black}{X-ray diffraction (XRD)} measurements, the spontaneous emission equations for the C and D transitions in one cavity alone are sufficient to solve directly for both the branching ratio and Purcell factors. However, by investigating a second, angled cavity, we both validate our model and demonstrate the advantages of aligning the cavity mode polarization as closely as possible to the dipole orientation of a specific transition. Indeed, if the angled cavity had been oriented $45\degree$ from the lattice vector rather than $55\degree$, from our analysis we would expect complete suppression of one transition and further increased enhancement of the other. However, by allowing for a controlled $10\degree$ of angular misalignment, we ensure that we are able to observe some degree of coupling between both C and D transitions of the emitter.

\section{Conclusion}\label{sec:conclusion}
In this manuscript we report the fabrication of 1D photonic crystal cavity nanobeams from thin film diamond membranes. We achieve upwards of $\sim6000$ quality factors, and observe up to 10-fold PL enhancement from a select SnV$^-$ when on resonance with the cavity mode. To accurately quantify our Purcell factors, we take time-resolved measurements and determine the optical lifetime reduction of the emitter. Despite optically filtering to isolate the C transition in collection, we also observe lifetime reduction when the cavity is on resonance with the D transition of the same emitter.

\textcolor{black}{From this picture, we construct a model describing the spontaneous emission dynamics of the system, with which we determine Purcell factors of $F_C=9.2 \pm 0.2$ and $F_D=8.9 \pm 0.1$ for the parallel cavity, and $F_C=26.2 \pm 1.5$ and $F_D=5.1 \pm 2.2$ for the angled cavity. Furthermore, by studying the cavity-emitter coupling behavior of these two devices we extract a C/D branching ratio of $\eta_\text{BR}\approx0.75\pm0.01$ and a collective fabrication angular offset of $\delta\psi\approx8.51\degree$.}

Our Purcell factor can be further increased through fabrication optimization, such as improving feature fidelity and minimizing device sidewall angles in the etch process (Appendix~\ref{app:cavity_yield}). Furthermore, cavity-SnV$^-$ coupling can be improved through targeted and aligned implantation of Sn$^{2+}$ into the cavity mode volume \cite{Sipahigil_2016}. Implantation density can be reduced for future devices in order to produce cavities with single color centers in the mode volume, while also reducing general implantation damage of the material. 

Our fabricated sample is compatible with microwave spin driving experiments. By enhancing the SnV$^-$ ZPL emission, readout fidelity of SnV$^-$ spins can be significantly enhanced ~\cite{Rosenthal_2023, Evans_2018}. Photon extraction can also be further improved by incorporating grating couplers or adiabatic tapers, which would enable color center addressing and photon collection in either transmission or reflection ~\cite{Burek_2017, Pasini_2024, Parker_2024, Nguyen_2019}. These developments would enable near-unity fidelity of single shot readout of the electron spin in the SnV$^{-}$, and pave the way for scalable quantum network nodes.

\begin{acknowledgments}
This work was supported by the Department of Energy, Grant No. DE-SC0025295. This work was partially supported by Center for Integrated Nanotechnologies (CINT), a DOE Office of Science user facility jointly operated by Los Alamos and Sandia National Laboratories. We acknowledge the U.S. Department of Energy via the Q-NEXT Center Grant No. DOE 1F-60579 and We acknowledge the use of shared facilities of the UCSB Quantum Foundry through Q-AMASE-i program (NSF DMR-1906325). H.C.K. acknowledges support by the Burt and Deedee McMurtry Stanford Graduate Fellowship (SGF). Work was performed in part in the nano@Stanford labs, which are supported by the National Science Foundation as part of the National Nanotechnology Coordinated Infrastructure under award ECCS-2026822. Part of this work was performed at the Stanford Nano Shared Facilities (SNSF), supported by the National Science Foundation under award ECCS-2026822. 

We would like to thank Lavendra Mandyam for helpful discussion in the cleanroom. 

\end{acknowledgments}

\appendix
\section{Thin Film and Device Fabrication}
\subsection{\label{app:thin_film_subsec}Diamond Thin Film Preparation}
The thin film diamond material used for these experiments was prepared using the `smart cut' technique and subsequent membrane exfoliation, as reported in \cite{Oh_2025, Lee_2013, Li_2023, Guo_2021}. 

Electronic grade, single crystal diamond material was sourced from Element 6 and polished by Syntek to a miscut angle of 1$\degree$. The polishing induced strained layer is removed via \textcolor{black}{reactive ion etching (RIE)} etching. A few microns is removed  primarily by Ar/Cl etch chemistry, followed by a short ($\sim$5 seconds) $\text{O}_2$ termination process to remove any Cl compounds on the surface. Helium ions (He$^{2+}$) were implanted into the diamond with an implantation energy of 150 keV at a fluence of $5 \times 10^{16}$ ions/cm$^2$, and the chip was subsequently annealed at 850 $\degree$C in a high-vacuum chamber. The graphite layer formed from He$^{2+}$ is $\sim$100 nm thick and $\sim$400 nm beneath the substrate surface \cite{Oh_2025}.

Due to the high implantation energy and dosage, the material above the graphitization is rendered unsuitable for downstream color center formation and device fabrication. A new layer of diamond is thus grown homoepitaxially on the top surface using plasma-enhanced chemical vapor deposition (PECVD). Detailed information on the growth procedure is reported in \cite{Hughes_2023}. 

To form tin vacancy centers, Sn$^{2+}$ ions were implanted by a commercial vendor, Cutting Edge Ions, into the newly grown diamond layer with an implantation energy of 380 keV and a fluence of $2 \times 10^{11}$ ions/cm$^2$. The Sn-implanted chip was high-vacuum annealed at 400$\degree$C for 4 hours, 800$\degree$C for 4 hours, and then 1200$\degree$C for 2 hours to promote color center formation. The sample was then tri-acid cleaned at 250$\degree$C for an hour to remove any surface graphitization that may have developed during the anneal. Lastly, a 90 minute UV ozone treatment oxygen terminates the sample surface.

\begin{figure}
    \includegraphics{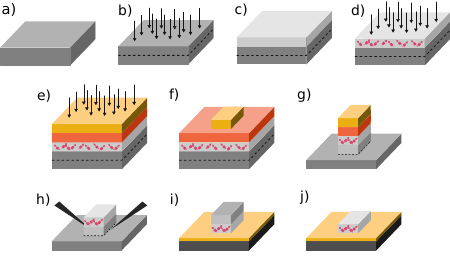}
    \caption{\label{sfig:thin_film_fab}Schematic of SnV$^-$ implanted thin-film diamond preparation. The process starts with a bulk electronic grade diamond that is polished to a precise miscut angle (a). The sample is then heavily implanted with He$^{2+}$ ions to form a vertically localized graphitized layer, around 400 nm below the sample surface (b). A pristine layer of single crystalline diamond is overgrown over the implantation damaged layer (c). Sn$^{2+}$ is implanted $\sim$90 nm below the sample surface, and SnV$^-$ centers are formed after high temperature, vacuum annealing (d). Any surface graphitization is removed via a tri-acid clean. A \textcolor{black}{silicon nitride (SiN)} hard mask is then deposited via \textcolor{black}{chemical vapor deposition (CVD)}, and spun with photoresist (e). 200 $\mu$m by 200 $\mu$m squares are patterned by photolithography; these squares define the final size of each membrane (f). The membranes are defined in the bulk diamond by an anisotropic RIE etch (g). Individual membranes are then released via electrochemical etch (h), and bonded to a Si carrier wafer with HSQ (i). Finally, the film thickness is tuned to 180 nm via RIE etching (j).}
\end{figure}

Membranes of size $200\times200\:\mu $m are patterned using photolithography and then released from the bulk using an electrochemical etch. The released membranes are then bonded damaged side up to Si carrier pieces via \textcolor{black}{the flowable oxide, hydrogen silsesquioxane (HSQ)}. Details on membrane exfoliation and transfer are outlined in \cite{Oh_2025}. The transferred membranes have a starting total thickness of ~$1\mu m$ and are thinned to the final desired thickness of 180 nm using \textcolor{black}{RIE}. We show a schematic of the full membrane preparation process in Fig.~\ref{sfig:thin_film_fab}.

\subsection{\label{app:cavity_fab_subsec}Photonic Crystal Fabrication Procedures}
\begin{figure}
    \includegraphics{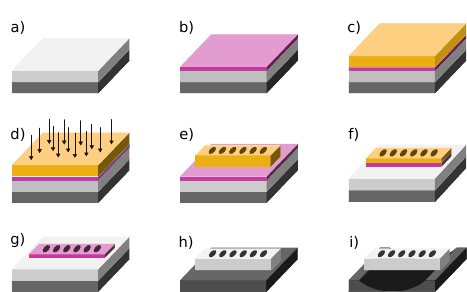}
    \caption{\label{sfig:cavity_fab} Schematic of cavity fabrication procedure. The starting material consists of thin film diamond bonded to a Si handling wafer via HSQ (a). A thin Al$_{2}$O$_{3}$ \textcolor{black}{hard} mask is deposited by ALD (b). The sample is spun with ZEP (c) and patterned via ebeam lithography (d, e). The resist pattern is transferred into the hardmask by ICP-RIE etching (f). Any remaining ZEP is then stripped (g), and the pattern transferred into the diamond membrane \textcolor{black}{(h)}. Lastly, the devices are suspended by HF vapor and XeF$_{2}$ dry etching (i).}
\end{figure}

Photonic crystal fabrication begins with the prepared 180 nm thin film membranes bonded to Si. A thin $\sim25$ nm $\textrm{Al}_2\textrm{O}_3$ hard mask is deposited via thermal \textcolor{black}{atomic layer deposition (ALD)}. The sample is spun with \textcolor{black}{a positive e-beam resist (ZEP 520A)} and then the pattern exposed via e-beam lithography. The resist pattern is transferred into the hardmask with a $\textrm{BCl}_3$ chemistry \textcolor{black}{inductively coupled plasma (ICP)} etch, and subsequently transferred into the diamond via a $\textrm{Cl}_2\textrm{O}_2 / \textrm{O}_2$ 2-step, cyclical ICP etch. The final devices are then released via HF and $\textrm{XeF}_2$ vapor dry etches. We show this process schematically in Fig.~\ref{sfig:cavity_fab}.

\subsection{\label{app:cavity_yield}Cavity Fabrication Yield and Fidelity}
\begin{figure*}
    \includegraphics{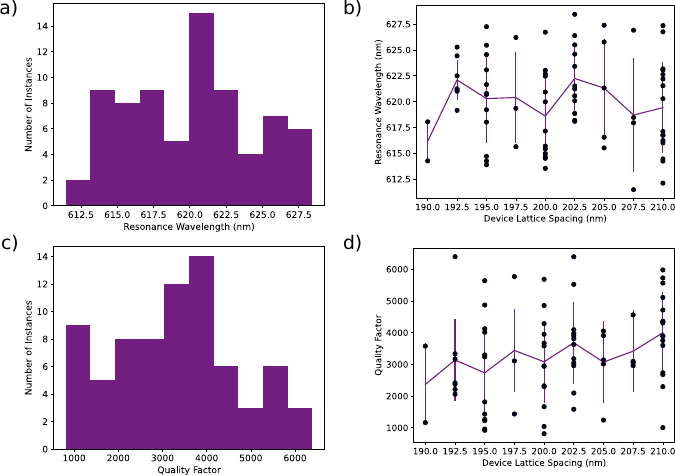}
    \caption{\label{sfig:cavity_yield}Summary of cavity fabrication yield across the sample.
    \textbf{(a)} Histogram of resonance wavelengths for devices. Across all identified resonances, the average resonance wavelength is 620$\pm 4$ nm.
    \textbf{(b)} Resonance wavelengths of all devices, summarized by the device lattice spacing. The black markers are data points from individual devices, while the purple line and errorbars indicate the average and standard deviation of of resonances wavelengths for devices of specific lattice spacings. From simulation we expect resonance wavelengths to red shift with increasing cavity lattice spacing; however, we do not observe a clear trend across our sample due to fabrication infidelities.
    \textbf{(c)} Histogram of quality factors for devices. Across all identified resonances, the average quality factor is 3335$\pm 1428$.
    \textbf{(d)} Quality factors of all devices, summarized by the device lattice spacing.The black markers are data points from individual devices, while the purple line and errorbars indicate the average and standard deviation of of quality factors for devices of specific lattice spacings. Quality factors are expected to remain roughly constant across all device designs. For our sample, quality factor very slightly increases with increased lattice spacings.}
\end{figure*}

To account for fabrication uncertainties, the cavity lattice spacing was swept from 180 nm to 210 nm, in steps of 2.5 nm, for a total of 13 device designs. All other design parameters (beam width, hole diameter, cavity lattice tapering factor) were kept constant. From FDTD simulations, we expect resonance wavelengths to correspondingly vary from $\sim$575 nm to $\sim$684 nm, in steps of $\sim$4 nm for each lattice spacing step.
In Fig~\ref{sfig:cavity_yield}, we survey all devices across our sample to evaluate our fabrication yield and fidelity. Due to fabrication variation, we do not observe a clear trend between device lattice spacing and resonance wavelength. Quality factors seem to demonstrate a very slight increase with increased lattice spacings. We also note that we were unable to find any resonances for devices with the lowest four lattice spacings, ranging from 180 nm to 187.5 nm, most likely due to the overetching of cavity holes. 

\begin{figure*}
    \includegraphics{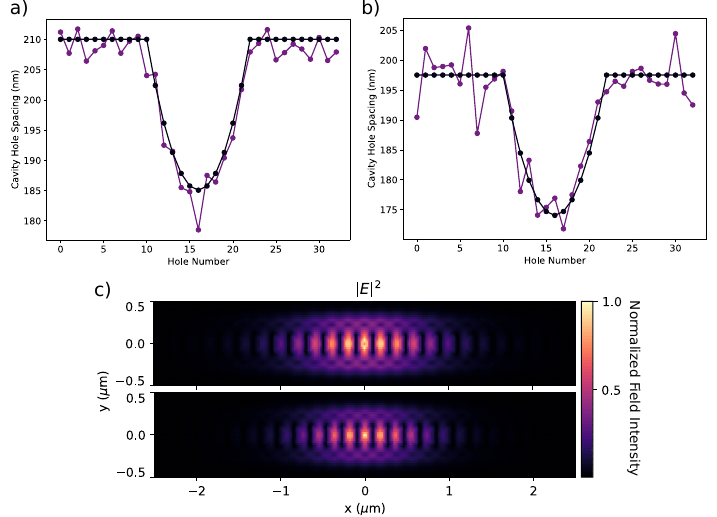}
    \caption{\label{sfig:cavity_fidelity} Comparison of cavity lattice spacings.
    \textbf{(a)} Lattice spacing deviations for the parallel device. On average the cavity holes deviated by $2.013 \pm 1$ nm from designed positions.
    \textbf{(b)} Lattice spacing deviations for the angled device. Cavity holes deviated by $2.796 \pm 3$ nm from designed positions.
    \textbf{(c)} Simulated normalized electric field inside the cavity region of the parallel device (top) and angled device (bottom).
    }
\end{figure*}

We evaluated the fabrication fidelity in detail for the two devices of study in the manuscript. Using Genisys ProSEM software, we analyze the tapered lattice spacings of the cavity holes (Fig.~\ref{sfig:cavity_fidelity}a,b). For the parallel device, we calculate a $2. \pm 1$ nm deviation from designed lattice spacings; for the angled device, we calculate $2.79 \pm 3$ nm deviation from designed spacings. Furthermore, we determine our device etch sidewall angle by fitting to both the top and bottom of the extracted beam width sigmoid. For the parallel device, we determine top (bottom) beam widths of $292.6 \pm 15$ nm ($358.6 \pm14$ nm); for the angled cavity, we extract $304.9 \pm 2$ nm ($337.3 \pm 13$ nm). Given a consistent membrane thickness of 180 nm, we  therefore determine sidewall angles of $10.389\degree$  and $5.143\degree$, respectively, which are included in the simulations. 

\textcolor{black}{Cavities were initially designed around and quality factors are maximized for vertical ($0\degree$) sidewalls. Therefore, any  sidewall angle resulting from fabrication serves as a primary source of quality factor degradation as trapezoidal waveguide cross-sections give rise to unwanted mixing between TE and TM modes. Furthermore, as Sn$^{2+}$ implantation depth targeted $90$ nm below the membrane surface, the diamond thin films were tuned to $180$ nm to best center the emitters in devices. However, quality factors may be boosted by thinning the membrane slightly further ($\approx20$ nm) in future iterations.}

\section{\label{app:measurement_setup}Characterization Setup}

\begin{figure}
    \includegraphics{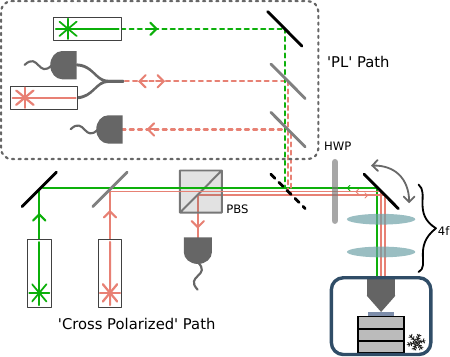}
    \caption{\label{sfig:measurement_setup}
    Schematic of the measurement setup. The optical setup primarily consists of a home built, 4f confocal microscope with scanning enabled by a galvo mirror (4f). A half wave plate (HWP) is inserted right before the galvo mirror, allowing for simultaneous polarization rotation of all excitation and collection paths. The optical setup incorporates two discrete optical paths, referred to in this manuscript as the 'PL' and the 'cross polarized' paths. Access to either of these two paths are controlled via a flip mirror (dashed diagonal line). In the cross polarized path, the excitation is horizontally polarized, and collection vertically polarized, with the two paths mixed via a \textcolor{black}{PBS}. Green diode and broadband supercontinuum excitation are separately fed into the optics setup, with the two paths mixed via a shortpass dichroic beamsplitter. For the PL path, collection is first split between green and red via a longpass diocrhoic beamsplitter, and then further split between ZPL and PSB via a shortpass dichroic beamsplitter. When addressing emitters resonantly, the resonant laser is launched from the ZPL collection fiber coupler.
    }
\end{figure}

Here we describe the optical setup used for device and emitter studies (Fig.~\ref{sfig:measurement_setup}). The optical setup consists of two separate access arms, referred to in the manuscript as the `PL' and `cross polarized' paths. Access between the two paths are controlled via a motorized mirror (Thorlabs, MFF101).

The sample is housed in a Montana Instruments Cryostation s50 at 5K, and positioned via a \textcolor{black}{triple} Attocube piezo stack (X101/Z100). The sample is addressed with a 100 $\mu$m working distance, 100x magnification objective (Zeiss, EC Epiplan-NEOFLUOR 100x/0.9), which is mounted in a heater cryo-objective housing. The 4f confocal scanning microscope is constructed via a pair of 300 mm lenses(Thorlabs, AC254-300-A) and a galvo mirror (Newport, FSM-300-01). Polarization control is provided by the HWP (Thorlabs, AHWP10M-600). 

In the cross polarized path, polarization filtering is provided by a polarizing beam splitter (Thorlabs, CCM1-PBS251). In the horizontally polarized collection path, the free space beam is coupled into a single mode (SM) fiber. In the horizontally polarized excitation arm, green and red wavelength splitting is achieved through a short-pass dichroic beamsplitter (Semrock, TSP01-561). The red excitation arm provides either resonant laser (Toptica, 1240 nm DL Pro, doubled via ADVR, frequency doubler) or broadband supercontinuum (superK, EVO EUL-10) light, coupled through a SM fiber. A Thorlabs, LP520-SF40 green diode laser is used for above band excitation. A HWP (Thorlabs, AHWP10M-600) and filter wheel is used to optimize and control power delivery, respectively. 

In the PL path, we split green and red with via a longpass \textcolor{black}{dichroic} beamsplitter (Semrock, DMLP550). Subsequently, we split ZPL and PSB via a shortpass dichroics (Semrock, FF625-SDi01). Pulsed, 520 nm excitation is provided by a Thorlabs, GSL52A laser. Either resonant excitation or signal can be launched or collected from the ZPL path. 

Collected signal is either routed to single photon counting modules (Excelitas, SPCM-AQRH-24) or a hybrid spectrometer (Acton, SpectraPro 2750 gratings and Andor, iDus416 CCD). For lifetime measurements, the parallel device lifetimes were collected on a Picoquant, Picoharp 300, while the angled device lifetimes were collected on a Swabian, Timetagger Ultra. Due to different timing resolution specs, the parallel device lifetimes were taken with 32 ps timing bins, and the angled device with 20 ps. During processing for the angled device, timing bins were further down-sampled to 40 ps. Pulse sequences are programmed and applied by a Swabian, Pulse Streamer 8/2. Pulsing and attenuation of the resonant excitation laser is provided by a G\&H, Fiber-Q 633 nm AOM. 

Fiber fluorescence from the green excitation sources are cleaned up with Thorlabs, FBH520-10 520/10 nm bandpass filters. Green scattering is further filtered from collection paths via Semrock, BLP01-594R 594 nm longpass filters. ZPL signal is isolated through Semrock, FBP01-620/14 620/14 nm bandpass filters. Filtering for isolating single SnV$^-$ transition lines was achieved using a custom ordered Rapid Spectral Solutions, 622/0.3 nm bandpass filter, designed to have $\sim$ 5 nm of tuning range. 

\section{Resonant Color Center Addressing}\label{app:resonant_PLE}

\begin{figure*}
    \includegraphics{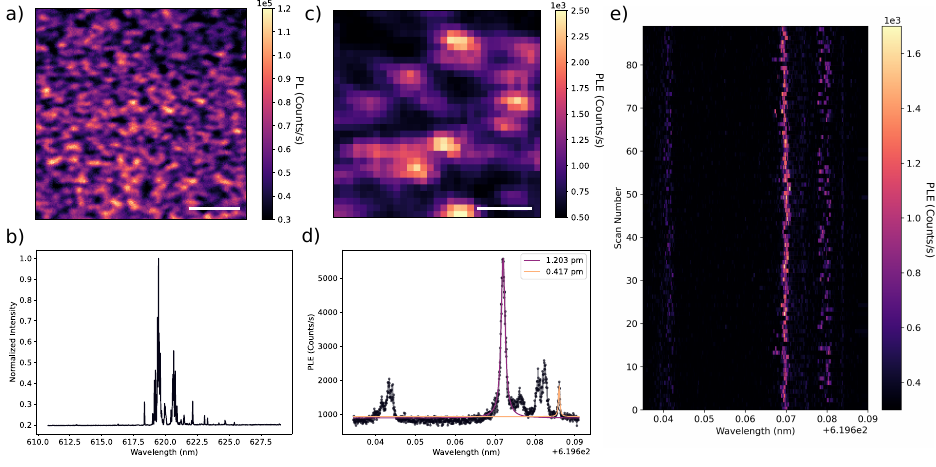}
    \caption{\label{sfig:PLE_prefab} Color center characterization in thin film, pre device fabrication.
    \textbf{(a)} PL confocal scan of the thin film pre device fabrication. The density of bright spots indicates the larger number of SnV$^-$s formed during implantation. The scale bar indicates 3$\mu$m.
    \textbf{(b)} Representative PL spectrum for a probed emitter cluster, demonstrating SnV$^-$ formation density.
    \textbf{(c)} PLE confocal scan. A full PLE scan is taken and count rates summed per pixel of the 2D spatial map. All emitters which show up are addressable within the laser scanning range of $\sim$40 GHz. The scale bar indicates 1$\mu$m.
    \textbf{(d)} Averaged PLE. Two representative transitions are fit with a Lorentzian model to estimate the linewidths. The more prominent transition demonstrates a 1.236 pm or $\sim$1 GHz linewidth. The other transition demonstrates a 0.417 pm or $\sim$320 MHz linewidth.
    \textbf{(e)} PLE traces over 30 minutes. The stability of the PLE indicates prolonged SnV$^-$ spectral coherence in the thin film.
    }
\end{figure*}

Prior to patterning photonic devices, optical characterization was performed to verify formation of color centers in the thin film material. Using 4f confocal microscopy, 2D PL spatial maps were obtained using an above-resonant (green, 520 nm) diode laser and bright spots probed with spectroscopy (Fig.~\ref{sfig:PLE_prefab}a). From Fig.~\ref{sfig:PLE_prefab}b, the large density of SnV$^-$s formed is evident from the PL spectra. This high emitter density is also evident when resonantly addressing the SnV$^-$ candidates. Within $\sim$40 GHz of laser scanning range, we can address on the order of 3-4 color centers (Fig.~\ref{sfig:PLE_prefab}d). 

To study the spectral coherence of these color centers, we perform PLE spectroscopy via the PL optical path. We excite the SnV$^-$ with alternating pulsed green and red excitation; SPCM collection is gated with red excitation to reduce signal background. The full pulse sequence is 20 $\mu$s, with 4 $\mu$s of green excitation, 4 $\mu$s of offset, and 12 $\mu$s of red excitation and SPCM collection. Pulsing for the green diode is provided by direct modulation of the laser diode, while pulsing for the red is provided by an AOM. The red laser is scanned across $\sim$40 GHz with a frequency of 0.2 Hz. We are also able to identify SnV$^-$ transitions with down to a $\sim$320 MHz linewidth, stable for $\sim$30 minutes (Fig.~\ref{sfig:PLE_prefab}e). However, color centers of higher coherence in both bulk and thin-film diamond have been reported \cite{Trusheim_2020,Guo_2023}. We attribute our measured linewidth to insufficient optimization of PLE experimental parameters, lack of an exhaustive survey of emitters, and the high Sn$^{2+}$implantation density. 

\begin{figure*}
    \includegraphics{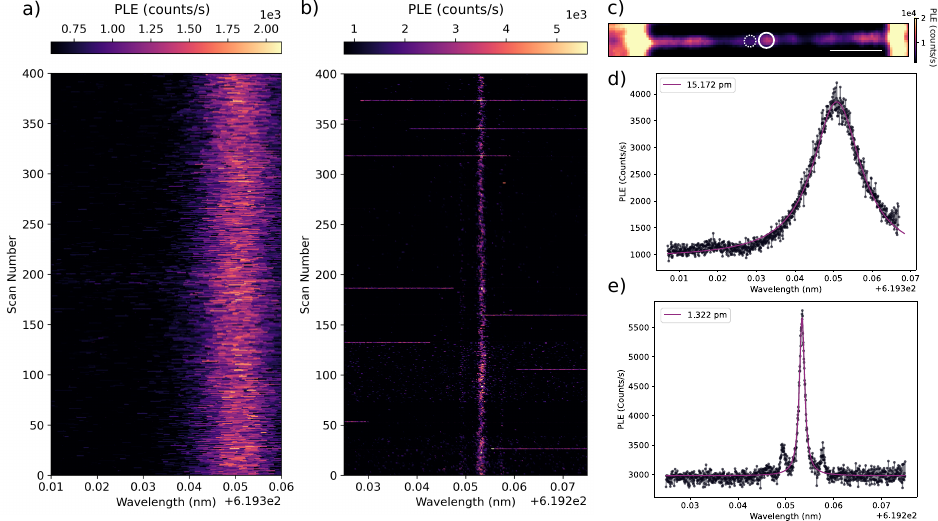}
    \caption{\label{sfig:PLE_postfab} Resonant color center addressing in fabricated devices.
    \textbf{(a)} PLE traces for $\sim$30 minutes on the color center of focus in the parallel device. 
    \textbf{(b)} PLE traces for $\sim$30 minutes for a secondary emitter in the parallel device. 
    \textbf{(c)} PLE confocal scan of the parallel device. The main emitter of study in this manuscript is indicated by the solid circle, while the dashed circle notates the secondary emitter of study.
    \textbf{(d)} Averaged PLE of the traces from (a). A Lorentzian fit yields a linewidth of 15.382 pm.
    \textbf{(e)} Averaged PLE of the traces from (b). A Lorentzian fit yields a 1.386 pm linewidth. We note that the side lobes of the main PLE peak feature was due to laser instability and multimodedness.
    }
\end{figure*}

Post-device fabrication, we first attempt PLE for the emitter of study in the parallel device. Although the PLE is consistent over $\sim$30 minutes, the emitter demonstrates significant linewidth broadening, with an averaged linewidth of $15.382$ pm (Fig.~\ref{sfig:PLE_postfab}a,d). We also probe a secondary emitter in the device which is much more weakly coupled to the cavity mode. This emitter instead demonstrates an averaged linewdith of $1.368$ pm; the PLE signal is similarly stable for over $\sim$30 minutes. The further linewidth broadening of the emitters in fabricated devices likely stemmed from the proximity of etched surfaces. To mitigate these effects, inspiration can be taken from efforts to improve near-surface NV coherence via surface passivation or chemical termination \cite{Sangtawesin_2019, Kumar_2024}.

\section{Extended Methods and Data}

\begin{figure*}
    \includegraphics{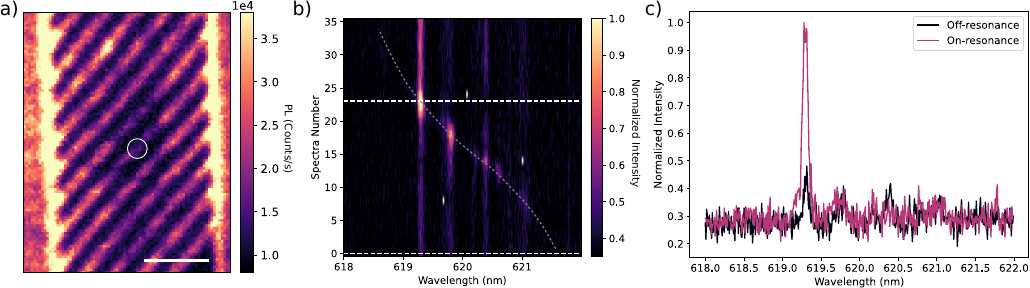}
    \caption{\label{sfig:angled_cavity_02}Angled cavity PL confocal scan and cavity enhancement.
    \textbf{(a)} PL confocal scan of the angled device studied in this manuscript. The emitter cluster of focus is indicated by the white circle. The scale bar indicates 3$\mu$m.
    \textbf{(b)} PL enhancement of the SnV$^-$s located in the angled device cavity mode. The spectra for the transition of interest on and off resonance are indicated by the white dahsed lines. The grey dashed line is a guide for the eye of the cavity resonance wavelength.
    \textbf{(c)} PL spectra of the two spectra of interest. From comparing the transition amplitude on and off resonance, we estimate a 2.5-fold PL enhancement from gas tuning.
    }
\end{figure*}

\subsection{\label{app:angled_cavity_2nd_emitter}Secondary Emitter in the Angled Cavity Mode}

\begin{figure*}
    \includegraphics{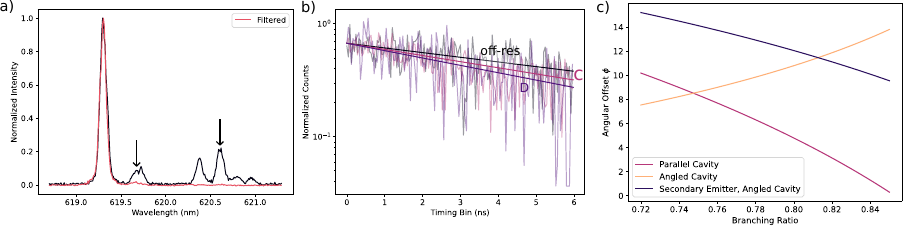}
    \caption{\label{sfig:angled_cavity_second_lifetime}Lifetime reduction of secondary emitter.
    \textbf{(a)} Filtered and unfiltered PL spectra of the collection spot for the angled device. Both are normalized in order to show degree of suppression of other transitions by the tunable filter. The arrows indicate the estimated C and D transitions of the second emitter. In the filtered spectrum, it is evident that a small contribution of the secondary C transition remains. 
    \textbf{(b)} Lifetime reduction for when the secondary emitter is on/ off resonance. The C transition demonstrates a reduced lifetime of $8.01 \pm 0.07$ ns, and D demonstrates a reduced lifetime of $6.66 \pm 0.20$ ns. The off resonance lifetime is $10.51 \pm 0.20$ ns.
    \textbf{(c)} Solution for Purcell analysis for three emitters. Although the three functions fail to intersect, minimizing the average difference yields $\eta_\text{BR}=0.75\pm0.06$, consistent with the value calculated in Section~\ref{subsec:purcell}. We also similarly determine the global lithographic angular offset to be$\delta\psi\approx8.51\degree$
    }
\end{figure*}

In this section we discuss the behavior of the second emitter in the angled cavity mode. Due to the bandwidth of the tunable optical filter ($\sim$0.3 nm), a small contribution from a second emitter is collected during lifetime measurements for the angled cavity. This contribution is visible when a PL spectrum is taken of the optically filtered collection spot with long integration times (Fig.~\ref{sfig:angled_cavity_second_lifetime}a).

When the C(D) transition is on resonance, the secondary emitter demonstrates a reduced lifetime of $8.01 \pm 0.07$ ns ($6.66 \pm 0.20$ ns). By fitting the data to the sum of four Lorentzians and extracting the relevant amplitudes, we determine amplitude ratios of \textcolor{black}{$\zeta_C=1.02 \pm 0.16$ and $\zeta_D=1.97 \pm 0.28$}. Given that the D transition demonstrates greater lifetime reduction than the C transition, we conclude that this secondary emitter is orthogonal in orientation to the other two SnV$^-$s of study. Using literature values for $\eta_{\text{DW}}$ and $\eta_{\text{BR}}$, we determine Purcell factors of \textcolor{black}{$F_C=1.1 \pm 0.4$ and $F_D=8.8 \pm 2.2$}, for the C and D transitions, respectively.

As performed in Subsection~\ref{sec:cavity_dipole_angle}, we can attempt to solve for $\eta_\text{BR}$ and $\delta\psi$, now taking into account this secondary emitter. From Fig.~\ref{sfig:angled_cavity_second_lifetime}c, we see that the three trend lines fail to fully intersect. This discrepancy is likely due to increased fitting errors, stemming from further reduced SNR per lifetime measurement for this more weakly coupled emitter. Therefore, we instead minimize the averaged pairwise differences between for all trend lines, identifying the optimal value \textcolor{black}{$\eta_\text{BR}\approx0.75\pm0.06$}, consistent with that determined in Section~\ref{subsec:purcell}. 

\subsection{\label{app:angled_cavity}Angled Cavity and PL Characterization}

\begin{figure*}
    \includegraphics{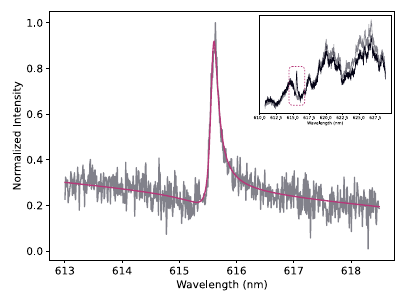}
    \caption{\label{sfig:angled_cavity_01}Angled cavity quality factor characterization. The full broadband reflectivity spectra for the resonance and background are shown in the inset. The fitted region is indicated by the dashed line. As with the parallel device, the resonance is fit to a Fano model. \textcolor{black}{In the inset, the background is represented by the grey trace, and the resonance signal by the black.}
    }
\end{figure*}

When measured in cross polarized reflectivity, the angled device yields a quality factor of $3942$, determined through a Fano fit on a background corrected broadband spectrum (Fig.~\ref{sfig:angled_cavity_01}). In Fig.~\ref{sfig:angled_cavity_02}, we also present the PL confocal scan of the angled device of interest. Similar to with the parallel cavity, the high implantation dose resulted in a cluster of 2-3 emitters coupled to the cavity mode. The most strongly enhanced transition demonstrates a $\sim$2.25-fold PL enhancement when on resonance with the cavity (Fig.~\ref{sfig:angled_cavity_02}). We note that this figure illustrates the necessity of lifetime reduction measurements for rigorously determining Purcell factors in a cavity QED system.

\subsection{\label{app:Q_determination}\textcolor{black}{Quality Factor Measurement}}

\begin{figure*}
    \includegraphics[width=0.65\textwidth]{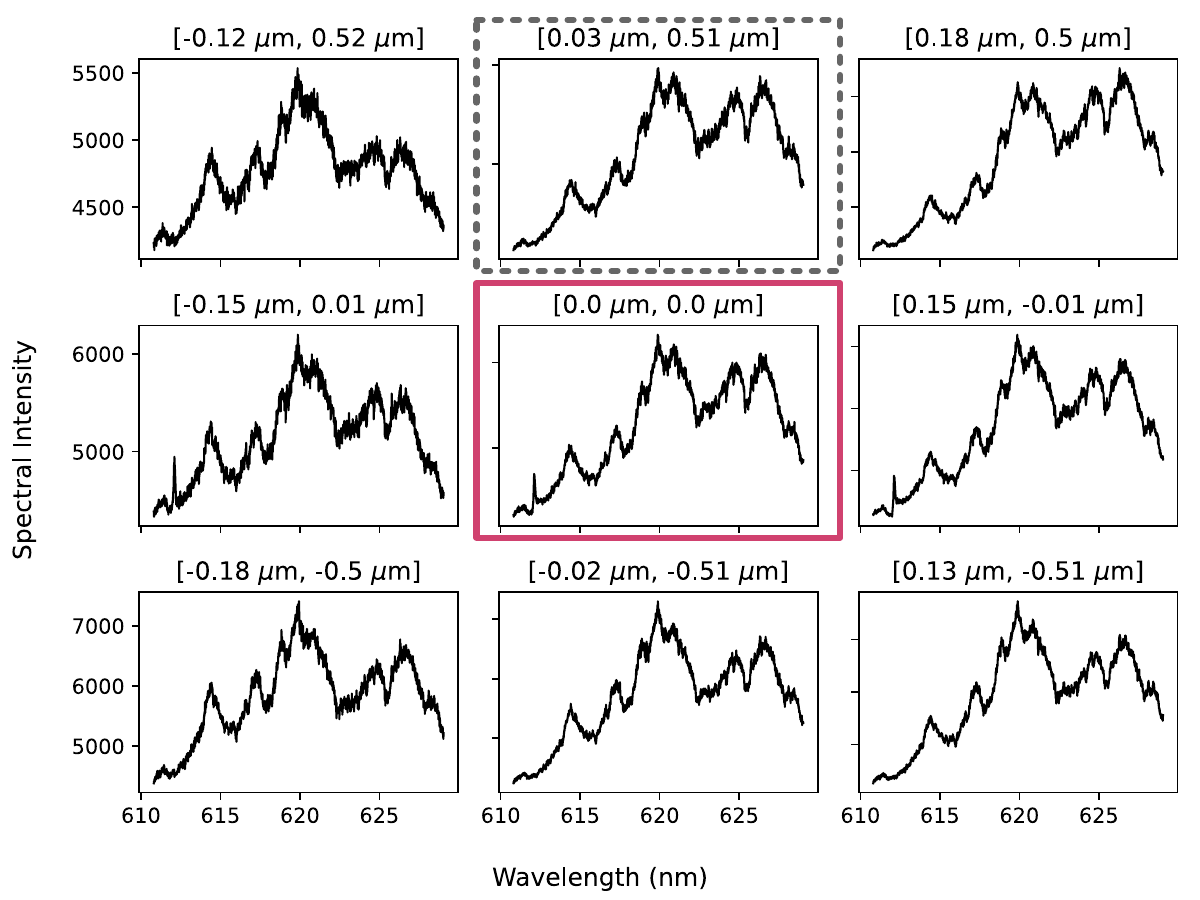}
    \caption{\label{sfig:spectra_location}\textcolor{black}{Example array of spectra from spatial location sweep to identify position of resonance for the parallel device presented in the main text. The spectrum used to measure the resonance is indicated by the pink box, and that used for background subtraction by the gray dashed box. Each spectrum is labeled with its spatial offset from the one used for determine the cavity quality factor.}
    }
\end{figure*}

\textcolor{black}{To probe for cavity resonances via cross-polarized reflectivity, we first generated a 2D map via PL confocal scanning, and then visually estimated the coordinates of the center of the photonic crystal cavity nanobeam. The galvo mirror was then used to perform a small spatial sweep around this estimated center (Fig.~\ref{sfig:spectra_location}). A broadband reflectivity spectrum was collected at each spatial location and spatially sensitive, narrow features were flagged as resonances. A spectrum that was located spatially adjacent to a location exhibiting a resonant feature was used for background subtraction.}

\begin{figure*}
    \includegraphics{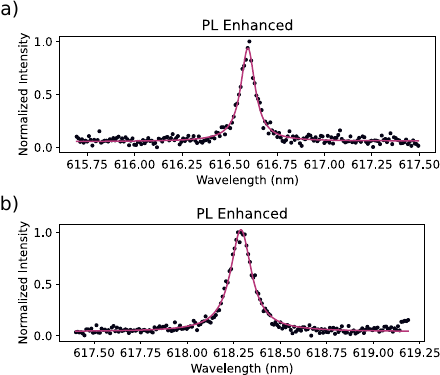}
    \caption{\label{sfig:q_determination} \textcolor{black}{Comparison of resonance fitting via PL spectra for both the parallel (a) and angled (b) devices. On this occasion, for the parallel (angled) device we measure a quality factor $6745\pm342$ ($4533\pm154$). From these fitted values, we conclude that the quality factors determined via cross-polarized reflectivity serve as lower-bound estimates.}
    }
\end{figure*}

\textcolor{black}{The cross-polarized reflectivity spectra are highly sensitive to small changes in HWP orientation, which was adjusted and optimized manually. This sensitivity, in conjunction with noise that arises from the reflection measurement (e.g. coupling of the excitation to free space modes above the light line), cause the resonances collected via this measurement technique to be Fano rather than Lorentzian in lineshape. This contributes an additional source of uncertainty in the extracted quality factors given the increased number of fitting parameters in the Fano model. To evaluate the degree of uncertainty in our quality factor determination, we compared the quality factors determined from cross-polarized reflectivity with those from PL excitation (Figure~\ref{sfig:q_determination}), wherein the cavity modes are excited via fluorescence from the color centers. When probing the quality factors via PL on the same devices reported in the main text, we determine a quality factor of $6745\pm342$ for the parallel device and $4533\pm154$ for the angled device. Both of these Q's were extracted from a Lorentzian model, which provides an appropriate fit for the resonance lineshapes measured via PL excitation. As expected, the quality factors derived from the Lorentzian-shaped resonances seen via PL excitation are higher than the quality factors extraced from the Fano-shaped reflectivity spectra. The Q's extracted from the reflectivity measurements are thus taken to be a lower bound.}


\subsection{\label{app:cavity_mode_pol}Cavity Mode Polarization Dependence}

\begin{figure}
    \includegraphics{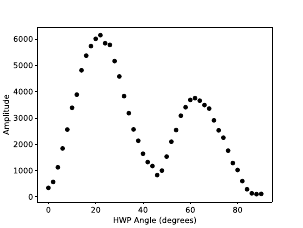}
    \caption{\label{sfig:pol_dep}Angular dependence of cavity reflectivity amplitude. The periodicity of the oscillatory behavior is $\sim$45$\degree$, indicating a linearly polarized mode.
    }
\end{figure}

To verify that the nanobeam cavity modes are linearly polarized, we measure the angular dependence of the reflectivity amplitude by rotating the HWP. The angular dependence demonstrates sinusoidal behavior with a periodicity of $\sim$45$\degree$, indicating a linearly polarized mode (Fig.~\ref{sfig:pol_dep})~\cite{Gong_2010}. The amplitude modulation of the sinusoidal trend is likely due to slight shifts in beam positioning introduced by HWP rotation.

\subsection{\label{app:lifetime_processing}\textcolor{black}{Lifetime Processing and IRF}}
\begin{figure}
    \includegraphics[width=\columnwidth]{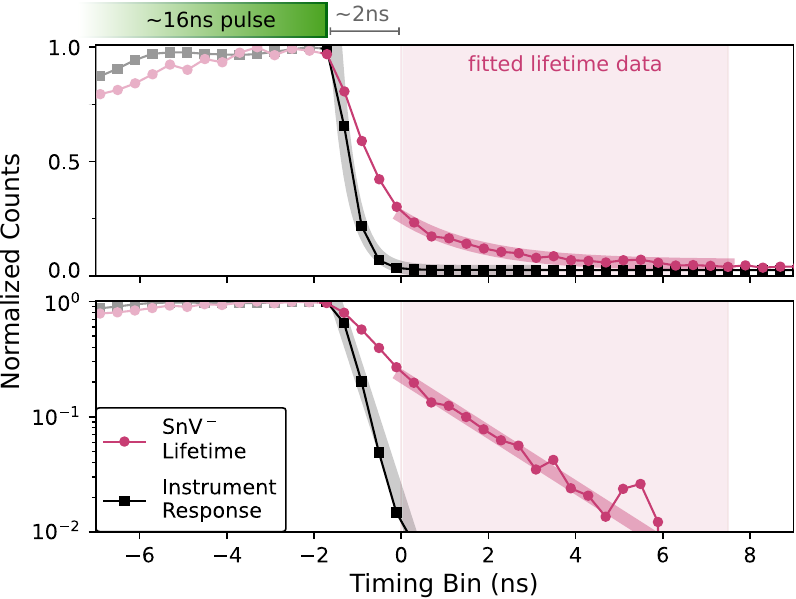}
    \caption{\label{sfig:IRF}\textcolor{black}{IRF of the optical setup. Color centers were excited with 16 ns pulses at a repetition rate of 3.3 MHz. The IRF is fitted to a single exponential model and yields a falltime of $\approx 370$ ps. Therefore, 2 ns of each lifetime trace is truncated, and the remaining data, shaded in pink, is fit to single exponential models. The most reduced lifetime trace, corresponding to $1.079$ ns is displayed in both linear (top) and logarithmic (bottom) scales to verify negligible effects from the IRF.}
    }
\end{figure}

\textcolor{black}{We characterize our optical setup's instrument response function (IRF) via the PL path. We focused the pulsed green laser on the sample surface, attenuated the power to prevent detector saturation, and then removed the longpass collection filter. The falltime of the IRF was then fit with a single exponential decay, yielding a falltime of $\approx370$ ps, slightly above the $300$ ps jitter expected of the SPCM detectors. To take this into account, the first $\approx2$ ns of each lifetime trace is therefore truncated and the remainder fit to a single exponential decay (Fig.~\ref{sfig:IRF}).}

\bibliography{bib}
\nocite{*}
\end{document}